\documentclass[fleqn,usenatbib]{mnras}

\usepackage{newtxtext,newtxmath}
\usepackage[T1]{fontenc}
\usepackage{ae,aecompl}

\usepackage{graphicx}	% Including figure files
\usepackage{amsmath}	% Advanced maths commands
\usepackage{amssymb}	% Extra maths symbols

\title[Revealing dust segregation with spectral index maps]{Revealing dust segregation in protoplanetary discs with the help of multi-frequency spectral index maps}

\author[Ya. Pavlyuchenkov et al.]{
Yaroslav Pavlyuchenkov,$^{1}$\thanks{E-mail: pavyar@inasan.ru}
Vitaly Akimkin,$^{1}$
Dmitri Wiebe,$^{1}$
Eduard Vorobyov$^{2,3}$
\\
$^{1}$Institute of Astronomy, Russian Academy of Sciences, Pyatnitskaya str. 48, Moscow, 119017, Russia\\
$^{2}$Institute of Physics, Southern Federal University,
Rostov-on-Don, 344090, Russia\\
$^{3}$University of Vienna, Department of Astrophysics, Vienna, 1180, Austria
}

\date{Accepted XXX. Received YYY; in original form ZZZ}

\pubyear{2019}

\begin{document}
\label{firstpage}
\pagerange{\pageref{firstpage}--\pageref{lastpage}}
\maketitle

\begin{abstract}
Dust is known to drift and grow in protoplanetary discs, which results in dust segregation over the disc extent. Maps of the spectral index $\alpha$ are a common tool for studying the dust content in protoplanetary discs. The analysis of observationally derived maps reveals significant gradients of the spectral index, confirming that dust evolves in the disc, but a more detailed information about the dust redistribution is required to make inferences about the early stages of dust growth. We calculated the spectral index maps based on the results of numerical hydrodynamical simulations using the {\tt FEOSAD} code, which allows studying a long-term dynamics of a self-gravitating viscous disc populated with coagulating, drifting, and fragmenting dust. Here we demonstrate that values of the spectral index estimated for different wavelength intervals within the far-infrared and radio bands reveal the presence of dust grains of various sizes. Specifically, we show that the disc regions with the maximal spectral index in a specific wavelength interval are the regions with the prevalence of dust grains of a specific size. Thus, a set of spectral index maps derived using different wavelength intervals can be used to recover the dust size-distribution over the disc extent.
\end{abstract}


%%%%%%%%%%%%%%%%%%%%%%%%%%%%%%%%%%%%%%%%%%%%%%%%%%

%%%%%%%%%%%%%%%%% BODY OF PAPER %%%%%%%%%%%%%%%%%%
\section{Introduction} \label{sec:intro}
One of the major goals in studying protoplanetary discs is to prove that they are indeed protoplanetary, that is to find some signatures of the planet formation. At later stages, when protoplanets are already formed, their presence can be inferred from some structural features in the disc (rings, spirals etc.). The earlier stages of planet formation should manifest themselves in some regular changes of dust properties, hinting at dust growth by coagulation.

A major observable that delivers an information on dust properties and distribution is a continuum spectrum in (sub-)millimetre range. For simplicity, this spectrum is often described by a single number, a spectral index $\alpha$, which characterizes a slope of the spectrum in these ranges. The value of $\alpha$ depends on dust optical properties and, thus, is assumed to be a tracer of, for example, variations in dust size distribution across the disc, related to grain growth, destruction, sedimentation, and size segregation. However, it is not easy to derive the information about the dust properties from the analysis of $\alpha$. The additional complexity is that the spectral index also depends on dust temperature and optical depth.

The spectral index is defined as 
\begin{equation}
\alpha=\frac{d\,\lg I_{\nu}}{d\,\lg \nu},
\end{equation}
where $I_{\nu}$ (erg cm$^{-2}$ s$^{-1}$ Hz$^{-1}$ sr$^{-1}$) is the radiation intensity at frequency  $\nu$. To deduce a relation between the value of $\alpha$ and dust properties, we begin with considering a case of emission produced by a uniform layer with constant physical parameters along the line of sight. In an optically thick medium, the emergent intensity is equal to the Planck function $B_{\nu}(T)$, where $T$ is the medium temperature. If either the frequency is low enough or the temperature is high enough to satisfy the Rayleigh-Jeans approximation ($x=h\nu/kT\ll1$), then $\alpha=2$ irrespective of the dust properties. 

In an optically thin medium, the emergent intensity is proportional to the Planck function multiplied by the dust optical depth $\tau_{\nu}$, i.e., $I_{\nu}=\tau_{\nu}B_{\nu}(T)$. The optical depth, in turn, is proportional to the dust absorption coefficient $\kappa_{\nu}$ (cm$^2$ g$^{-1}$) and dust surface density $\Sigma_{\rm dust}$, i.e., $\tau_{\nu}=\kappa_{\nu}\Sigma_{\rm dust}$. In the far-infrared, the absorption coefficient is usually parameterized by a power-law $\kappa_{\nu}=\kappa_0(\nu/\nu_0)^{\beta}$ with $\beta$ being the opacity index:
\begin{equation}
\beta=\frac{d\,\lg \kappa_{\nu}}{d\,\lg \nu}.
\end{equation}
In the optically thin limit and the Rayleigh-Jeans regime, the spectral index is explicitly related to the opacity index as $\alpha=2+\beta$. As $\beta$ is mainly defined by the properties of the grain size distribution, spatially resolved maps of $\alpha$ can in principle be used to infer the dust size segregation in protoplanetary discs. Here and below by a dust size we mean a grain radius.

However, in other cases, in particular when $\tau_{\nu} \approx 1$ and $x\gtrsim1$, the situation is more complicated. In this case, the emergent intensity $I_{\nu}=B_{\nu}(T) (1-e^{-\tau_{\nu}})$. Differentiating this expression over frequency, we obtain a more general formula for $\alpha$:
\begin{equation}\label{eq:main}
\alpha=3-\frac{x}{1-e^{-x}}+\frac{\tau_{\nu}}{e^{\tau_{\nu}}-1}\beta.
\end{equation}
The first two terms on the right-hand side of this expression reflect the spectral slope of the Planck function $\alpha_{\rm P}=d\lg B_{\nu}/d\lg \nu=3-x/(1-e^{-x})$. Obviously, $\alpha_{\rm P}=2$ in the Rayleigh-Jeans domain ($x\ll1$), while in the Wien limit ($x\gg1$) the spectral slope $\alpha_{\rm P}=3-x$ can even be negative. The factor $\tau_{\nu}/(e^{\tau_{\nu}}-1)$ accounts for the optical depth effects\footnote{Note that this factor differs from the classical $\Delta$-correction proposed in~\citet{1990AJ.....99..924B} for spatially unresolved observations.}. 

Since the spectral index is in general a function not only of $\beta$ but also of the optical depth $\tau$ and temperature $T$, deriving dust properties from the maps of $\alpha$ can be far from trivial. For example, $\alpha=2$ may imply both an optically thick layer with `typical' dust properties and an optically thin layer, in which dust opacity is dominated by large dust grains ($\beta=0$).

The spectral index and the corresponding opacity indices have been adopted by many authors as basic tools to study dust in protoplanetary discs, see, e.g., \citet{2003A&A...403..323T, 2006A&A...446..211R,2010ApJ...714.1746I, 2010A&A...521A..66R, 2010A&A...512A..15R, 2011A&A...525A..12B, 2012MNRAS.425.3137U} and many others. The general outcome of these studies is that protoplanetary discs at sub(mm) wavelengths reveal low spectral indices $\alpha\approx 2.5$ (the corresponding opacity index is $\beta\approx 0.5$ in the optically thin approximation), indicating that dust has been reprocessed and has grown up to a millimetre size. Moreover, high angular resolution images of protoplanetary discs at mm and sub-mm wavelengths \citep[e.g.][]{2012ApJ...760L..17P,2015ApJ...813...41P,2016A&A...588A..53T,2016ApJ...816...25P} inferred a radial gradient of the spectral index, pointing to the redistribution of large grains toward the disc center or prominent particle growth in the inner disc regions. At the same time, a variation of $\alpha$ is detected in discs with a resolved ring structure, indicating the dust segregation between the rings and gaps \citep[see, e.g.][]{2016ApJ...829L..35T, 2018ApJ...852..122H}. Extensive observational studies of dust in protoplanetary discs have been accompanied by a number of theoretical works on spectral and dust opacity indicies. For instance, \citet{2010A&A...516L..14B} studied how the spectral slopes at (sub-)mm wavelength depend on the details of dust coagulation/fragmentation using a dust evolution model coupled to a disc-structure model. In their recent study, \citet{2018arXiv181204043B} showed how steady-state dust size distributions in a coagulation-fragmentation equilibrium affect dust opacities by introducing dependencies on temperature, surface density, turbulence, and material properties. 

In our study, we calculate the spatial distributions of the spectral index based on a realistic numerical hydrodynamical model of evolving protoplanetary disc with coagulating dust, and show how these maps can be used to reveal dust evolution in a disc.

\section{Spectral index charts} \label{sec:spopin}

\begin{figure}
\includegraphics[width=\columnwidth]{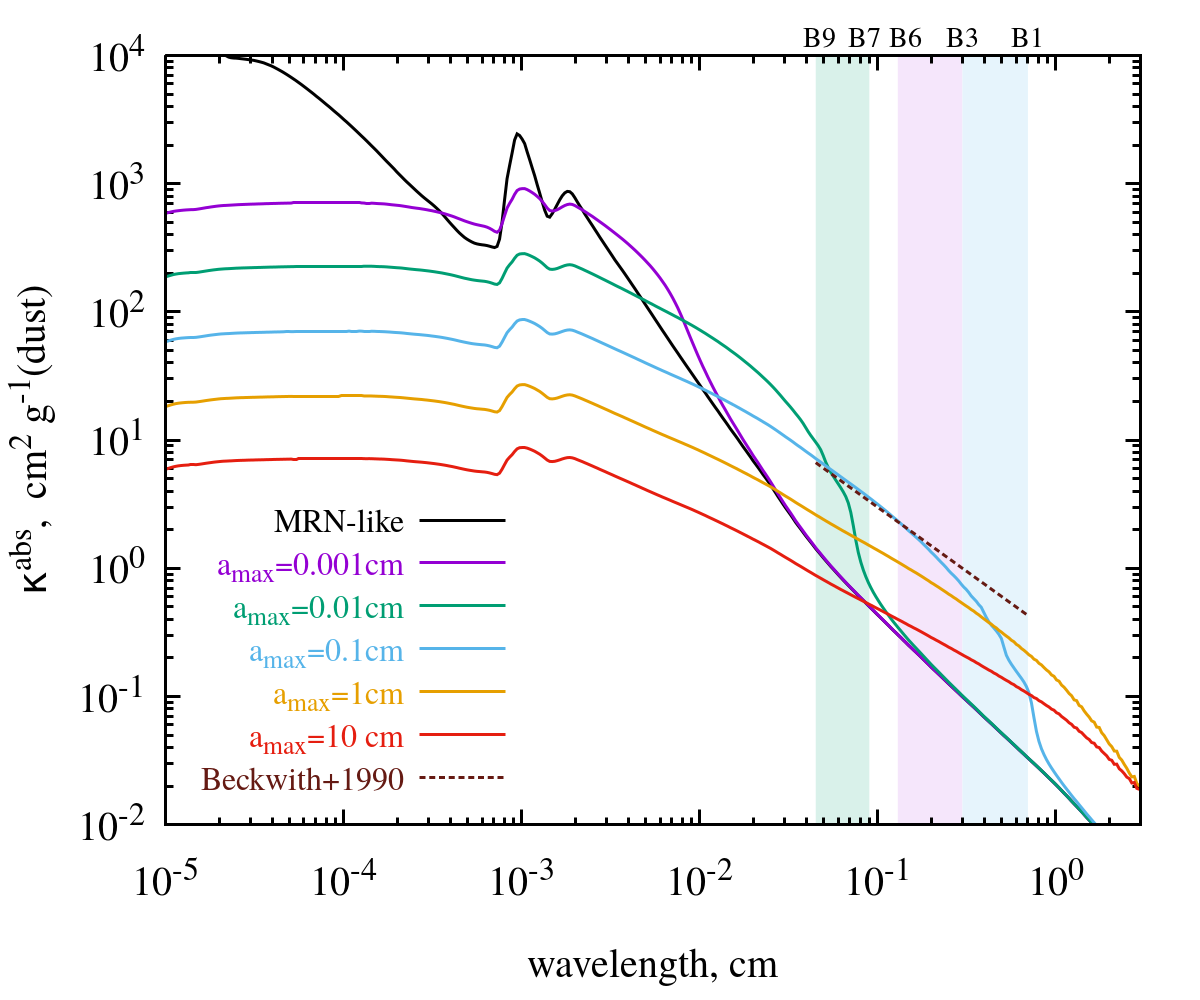}
%\fig{fig01.png}{0.45\textwidth}{}
\caption{Frequency dependence of dust absorption coefficient for different choices of the maximum grain size. Black solid line represents the `small dust' population for which we assume $a_{\rm min}=0.005\,\mu$m and $a_{\rm max}=1\,\mu$m. The solid coloured lines correspond to the `grown dust' population with fixed $a_{\rm min}=1\,\mu$m and varying $a_{\rm max}$ as shown in the legend. The power-law slope in the dust size distribution for small and grown dust is fixed at $p=-3.5$. The dashed line shows opacities from \citet{1990AJ.....99..924B} $\kappa_{\nu}=10(\nu/10^{12}\mbox{Hz})^{\beta}$~cm$^2$~g$^{-1}$ with\, $\beta=1$. The vertical colour bands correspond to the wavelength intervals over which we calculate $\alpha$. The boundaries of these intervals lie withing ALMA bands which are labeled above the graph.}
\label{fig1}
\end{figure}

The relation between the opacity index and dust properties plays a crucial role in the interpretation of observed intensity maps. A common wisdom states that grain growth manifests itself in decreasing values of $\beta$, but this assumption is, in fact, over-simplified. To draw a more general picture, let us consider how the dust absorption coefficient $\kappa_{\nu}$ changes with dust parameters in the framework of a particular dust model. For our analysis, we use a mixture of silicate and carbonaceous grains with the mass ratio of 0.8:0.2, which agrees reasonably  with estimates in \citet[][pp.265--266]{2011piim.book.....D}. The grain size distribution is assumed to be an MRN-like power law with a fixed exponent $p=-3.5$ \citep{1977ApJ...217..425M}. The minimum and maximum grain sizes $a_{\rm min}$ and $a_{\rm max}$ are treated as parameters. Refraction indices for silicate grains represent the so-called `astronomical silicates' from \citet{dl1984}\footnote{\url{https://www.astro.princeton.edu/\~draine/dust/dust.diel.html}}. The corresponding data for carbonaceous grains are taken from the Database of Optical Constants for Cosmic Dust, Laboratory Astrophysics Group of the AIU Jena\footnote{\url{https://www.astro.uni-jena.de/Laboratory/OCDB/index.html}}. Specifically, we use the data obtained from a pyrolysis experiment at $400^\circ$C \citep{jager1998}. The Mie theory is used to compute the corresponding opacities. In Fig.~\ref{fig1} we show the wavelength dependent opacity profiles, $\kappa_{\nu}$, for various maximum grain sizes. One can see that variations in $a_{\rm max}$ lead to changes in both opacity value and opacity slope.

\begin{figure}
\includegraphics[width=\columnwidth]{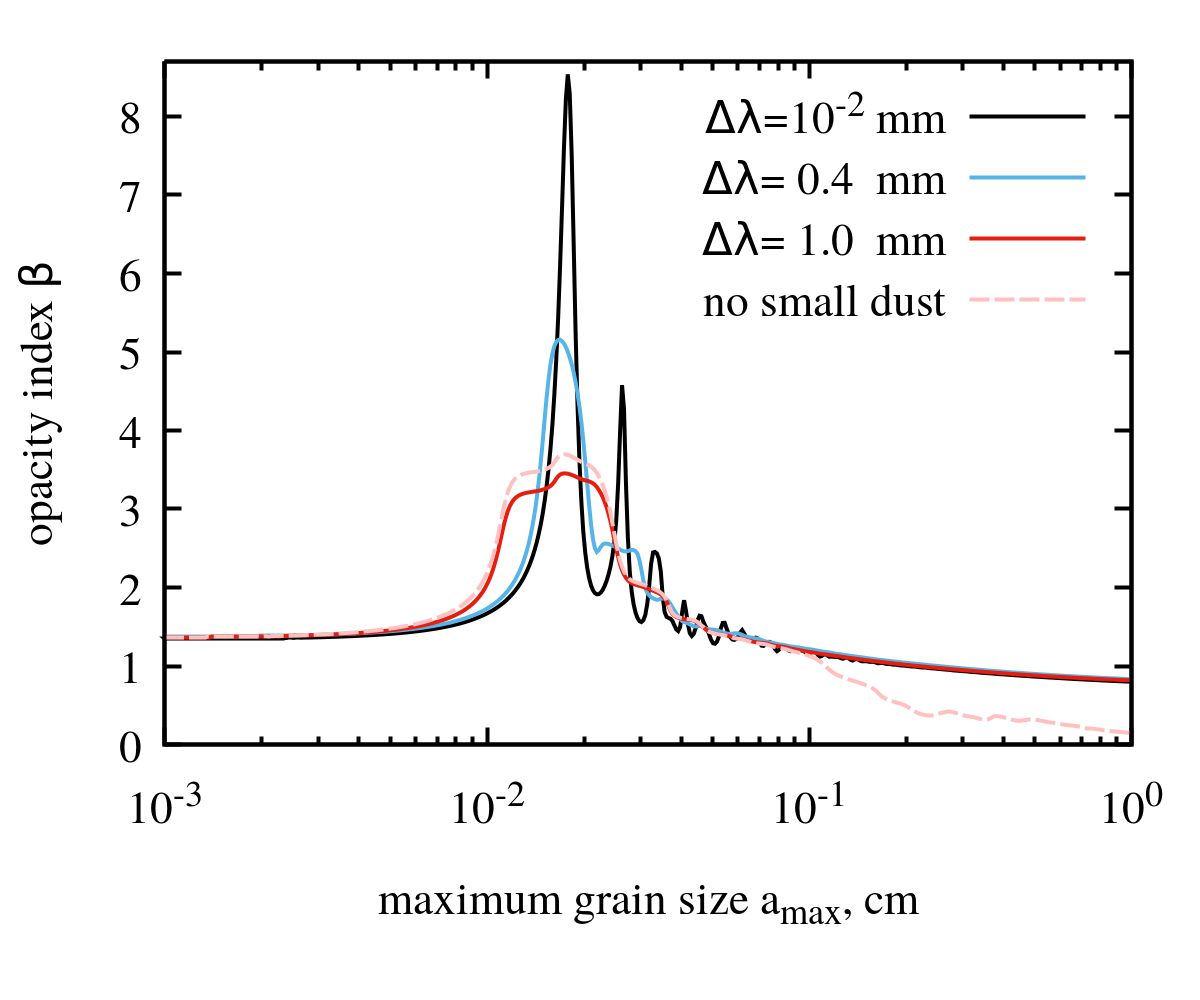}
%\fig{beta_width.png}{0.45\textwidth}{}
\caption{Dependence of the opacity index $\beta$ on the maximum grain size $a_{\rm max}$ for different wavelength intervals used to calculate the index. The width of the interval $\Delta \lambda=\lambda_2-\lambda_1$ is indicated in the legend, the interval is centered at $(\lambda_1+\lambda_2)/2=1.3$\,mm. The black line for $\Delta \lambda=10^{-2}$\,mm is very close to the genuine behaviour, while wider intervals (blue and red lines) smear it out. The minimum grain size for solid lines is $a_{\rm min}=1\,\mu$m. The pink line corresponds to $\Delta\lambda=1.0$\,mm and $a_{\rm min}=a_{\rm max}/10$, i.e., to the model without small grains. Note that near-zero values of $\beta$ require not only the presence of large dust, but also the absence of small dust.} \label{fig2}
\end{figure}

Due to the noise in observational data, the spectral and opacity indices are often derived using a relatively broad wavelength interval $(\lambda_1,\lambda_2)$:
\begin{eqnarray}
&&\alpha(\lambda_1,\lambda_2) = 
\frac{\lg I(\nu_1)-\lg I(\nu_2)}{\lg \nu_1-\lg \nu_2} \label{eq_alpha_fin}\\
&&\beta(\lambda_1,\lambda_2) = 
\frac{\lg \kappa(\nu_1)-\lg \kappa(\nu_2)}{\lg \nu_1-\lg \nu_2}.\label{eq_beta_fin}
\end{eqnarray}
The specific choice of the interval width affects the inferred dependence of the opacity index $\beta$ on the maximum grain size which is illustrated in Fig.~\ref{fig2}. The black curve corresponds to a nearly ideal situation, when the interval width is relatively small ($\Delta \lambda=\lambda_2-\lambda_1=10^{-2}$\,mm). At $a_{\rm max}<5\cdot 10^{-3}$~cm, the opacity index profile is nearly flat with a value around 1.4, then several prominent peaks appear, and after $a_{\rm max}=5\cdot 10^{-2}$ the opacity index slowly decreases with growing $a_{\rm max}$. The set of local peaks on the $\beta(a_{\rm max})$ plot is localized around $a_{\rm max}\sim \lambda/2\pi$ (the exact positions of maxima depend on the grain optical properties, the grain size distribution, the strength of the interference and the ripple structures, for more details see section 4.4.2 in \citet{1983asls.book.....B}). As we increase the interval width $\Delta \lambda$, the $\beta(a_{\rm max})$-dependence transforms into a single peak, and its width also grows. The similar $\beta$-profiles have been presented in a number of studies \citep[see e.g.][]{1994ApJ...421..615P, 2004A&A...416..179N, 2014prpl.conf..339T, 2015PASP..127..961A, 2018arXiv181204043B}.

It is worth noting that the validity of the conventional statement on the association of grain growth with the zero opacity index crucially depends on the details of the grain growth. As long as the lower limit $a_{\rm min}$ of the dust size distribution is fixed, $\beta$ does not go to zero even at large $a_{\rm max}$ (see Fig.~\ref{fig2}). On the other hand, opacity index does go to zero, if an increase of $a_{\rm max}$ is accompanied by a simultaneous increase of $a_{\rm min}$. An example is shown in Fig.~\ref{fig2} with the pink dashed line, which corresponds to the case when $a_{\rm min}$ is not fixed but is taken to be one tenth of $a_{\rm max}$.

\begin{figure*}
\centering
\includegraphics[width=\columnwidth]{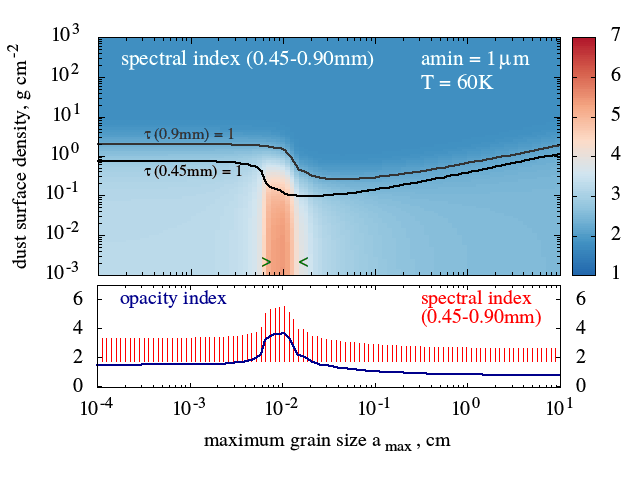}
\includegraphics[width=\columnwidth]{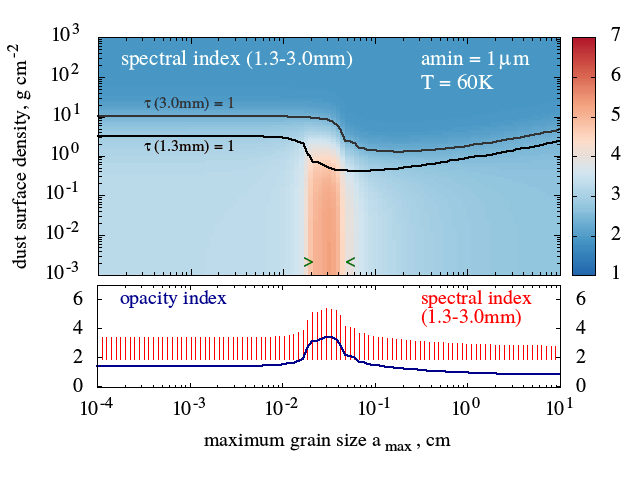}
\includegraphics[width=\columnwidth]{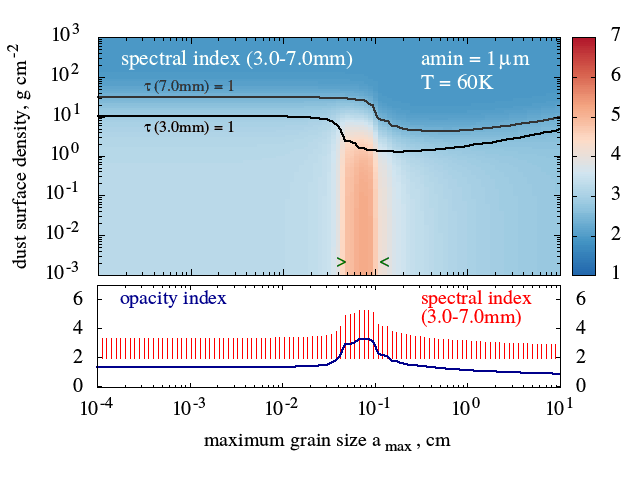}
\includegraphics[width=\columnwidth]{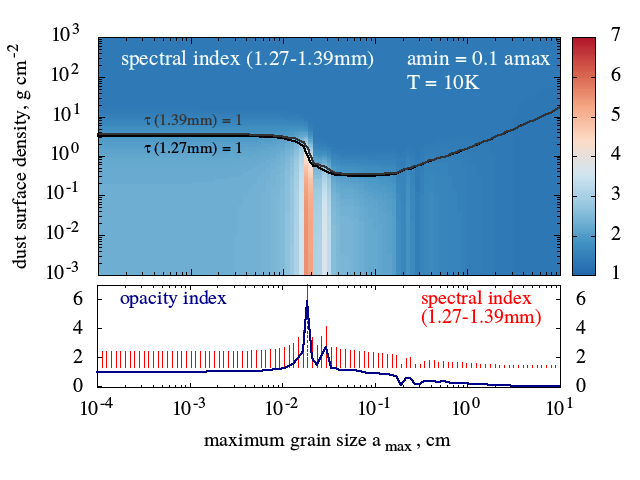}
%\fig{fig02a.png}{0.45\textwidth}{}
%\fig{fig02b.png}{0.45\textwidth}{}\\
%\fig{fig02c.png}{0.45\textwidth}{}
%\fig{fig02d.png}{0.45\textwidth}{}\\
\caption{Spectral index charts for the selected intervals. Top left: 0.45--0.90~mm (B9--B7 ALMA bands). Top right: 1.3--3.0~mm (B6--B3). Bottom left: 3.0--7.0~mm (B3--B1). Bottom right: 1.27--1.39~mm (within B3). First three charts are calculated using $T=60$~K and a power law size distribution with $a_{\rm min}=1\,\mu$m, while the last one corresponds to $T=10$~K and $a_{\rm min}=a_{\rm max}/10$. Each chart consists of an upper and lower panel. Upper panel: spectral index as a 2D function of the maximum grain size $a_{\rm max}$ and dust surface density. Green angle brackets indicate the interval range $\lambda_1/2\pi<a_{\rm gr}<\lambda_2/2\pi$. Lower panel: the dependence of the opacity index on the maximum grain size (blue). The red vertical lines show the extent of the spectral index $\alpha$ at a given $a_{\rm max}$ derived using all the values of $\Sigma_{\rm dust}$ from the above panel. 
\label{fig3}}
\end{figure*}

A more refined view is presented in Fig.~\ref{fig3}, where we show the spectral index charts calculated using a one-layer model. Namely, we represent the medium by a homogeneous layer with predefined temperature $T$ and dust surface density $\Sigma_{\rm dust}$. The emergent intensity $I_{\nu}$ is given by 
\begin{equation}
I_{\nu} = B_{\nu}(T)\left(1-\exp(-\tau_{\nu})\right),
\label{eq_RT}
\end{equation}
where $\tau_{\nu}=\kappa_{\nu}\Sigma_{\rm dust}$ is the optical depth, and $\kappa_{\nu}$ is the dust opacity calculated for given $p$, $a_{\rm min}$ and $a_{\rm max}$. The emergent spectral index and opacity index are calculated with Eqs.~\eqref{eq_alpha_fin} and~\eqref{eq_beta_fin}, respectively. The spectral index is produced for three wavelength intervals: 0.45--0.90~mm, 1.3--3.0~mm, and 3.0--7.0~mm (which are indicated in Fig.~\ref{fig1} with vertical colour bands). The interval boundaries are chosen to differ by about a factor of two to make the intervals relatively broad. At the same time, the boundaries of intervals lie within ALMA bands, i.e. we define the spectral index between the bands B9 -- B7, B6 -- B3, and B3 -- B1\footnote{https://www.eso.org/public/teles-instr/alma/receiver-bands/}. These three charts are calculated assuming a temperature of $T=60$~K and a power-law dust size distribution $n(a) \sim a^{-3.5}$ with fixed $a_{\rm min}=1\,\mu$m. Each chart consists of two panels. In the upper panel, we show the spectral index as a 2D function of the maximum grain size $a_{\rm max}$ (horizontal axis) and dust surface density (which defines the optical depth) in the considered plane-parallel layer (vertical axis). In the lower panel, the corresponding profile of the opacity index (blue lines) is presented together with the set of the spectral index profiles stacking from the above panel for all the values of $\Sigma_{\rm dust}$ (red hatched areas). 

A prominent feature in each of these charts is a region with elevated spectral indices in the optically thin part of the map, which stands out by its reddish colour. The position of this region is quite well constrained by the dust size interval $\lambda_1/2\pi<a_{\rm max}<\lambda_2/2\pi$, where $\lambda_1$ and $\lambda_2$ are the boundaries of wavelength interval used to evaluate $\alpha$ (see green angle brackets in the top panels). Obviously, the peak position shifts with changing the wavelength interval. As expected, $\alpha=2$ in the optically thick regime, which lies above the black curves delineating an optical depth of 1.0. In the optically thin case, $\alpha$ is nearly constant outside the red regions, being in average slightly higher at smaller $a_{\rm max}$. We note that $\alpha$ does not converge to a canonical value of 2 for high $a_{\rm max}$ because of the presence of small grains in the adopted dust size distribution.

To demonstrate the sensitivity of the spectral index to other parameters, in the right bottom panel of Fig.~\ref{fig3} we also present the spectral index chart calculated for a much narrower wavelength interval 1.27--1.39~mm, lower temperature $T=10$~K, and a truncated dust size distribution with $a_{\rm min}=0.1\,a_{\rm max}$. A peak of large $\alpha$ in a lower subpanel becomes higher and narrower and is also supplemented by several sub-peaks to its right. Since at such a low temperature the Rayleigh-Jeans approximation fails, the minimum value of $\alpha \approx 1.8$, that is, lower than 2. Another important difference from the other charts is that $\alpha$ converges to the minimum value of $\approx 1.8$ in the optically thin part at $a_{\rm max}>1$~cm. This is the result of not only the presence of large grains but also the absence of small grains in the adopted dust size distribution.

To summarize this section, we emphasize that the most prominent feature of the considered spectral index distributions
is the occurrence of strong maxima in $\alpha$, which are related to maximal dust sizes of the corresponding dust
size distributions. As we will see later, this correlation can be used to pinpoint the locations of dust grains with particular sizes in a protoplanetary disc.

We note, however, that the strong peak in $\beta$ profile is critically related to the adopted dust model, namely, to the representation of the dust as an ensemble of compact spherical grains. Meanwhile, dust coagulation in protoplanetary discs may also produce  aggregates with significant porosity, namely with filling factors ranging from 10$^{-4}$ to 1, see e.g. \citet{2008ApJ...684.1310S,2008ApJ...677.1296W, 2012ApJ...752..106O}. Such grains have optical properties that are very different from those of compact grains used here. Most notably, the main peak in the $\beta$ versus $a_{\rm max}$ profile (see Fig.~\ref{fig2})\, disappears for porous aggregates (see Fig.~11b in \citet{2014A&A...568A..42K}). Other factors that can potentionally smooth out the peak in the $\beta$-profile are the non-spherical shape of the dust grains\footnote{\url{http://www.astro.spbu.ru/DOP/8-GLIB/BASICS/index.html}} and grain chemical composition \citep{1994ApJ...421..615P}. Keeping all these factors in mind, we explicitly state that the results presented below apply the adopted dust model of compact homogeneous spherical grains.

\section{Spectral index maps for a circumstellar disc model with dust evolution} \label{sec:diskres}

In this section, we illustrate how the spectral index maps can be used to study the dust content in protoplanetary discs based on the charts presented in Fig.~\ref{fig3}. We use the {\tt FEOSAD} numerical hydrodynamics code described in \citet{2018A&A...614A..98V} to model the long-term evolution of a circumstellar disc starting from its formation and ending in the T Tauri phase. The evolution of gas and dust is computed self-consistently with the dynamical, thermal, and gravitational processes taken into account. The model includes a dust component consisting of two families: small sub-micron-sized grains and grown dust with a maximum grain radius $a_r$. The former is assumed to be strictly coupled to the gas, while the latter can drift through the gas. Grain growth and their destruction, as well as conversion of small to grown dust, the back reaction of dust on gas, and self-gravity of dust are considered. Numerical hydrodynamics simulations start form a pre-stellar core with a mass of 1.03~$M_\odot$ and a ratio of rotational to gravitational energy of $\beta=0.24\%$. The initial dust-to gas ratio is set to a canonical value of 0.01 and all dust is initially assumed to be in the form of small sub-micron-sized grains. The initial temperature of the pre-stellar core is set equal to 20~K. The initial pre-stellar core is similar to the  Bonnor-Ebert sphere with a central plateau and gas surface density declining at the core periphery. For more details on the initial configuration we refer to \citet{2018A&A...614A..98V}.

Using the distributions of dust density, maximum dust radius and dust temperature obtained by the {\tt FEOSAD} code,  we calculated the radiation intensity distributions for the wavelengths that were selected  to produce spectral index maps. The intensities were calculated using Eq.~\eqref{eq_RT} assuming that the disc is viewed face-on and that the temperature does not change along the vertical direction. Based on the calculated intensities, we produced the spectral index maps using Eq.~\eqref{eq_alpha_fin}. We also calculated the maps of optical depth for the sake of analysis.

\begin{figure*}
\centering
\includegraphics[width=\columnwidth]{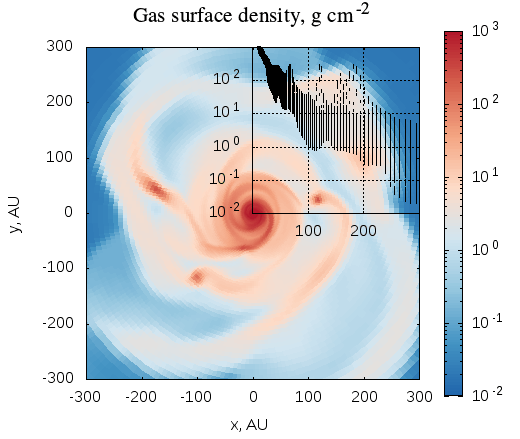}
\includegraphics[width=\columnwidth]{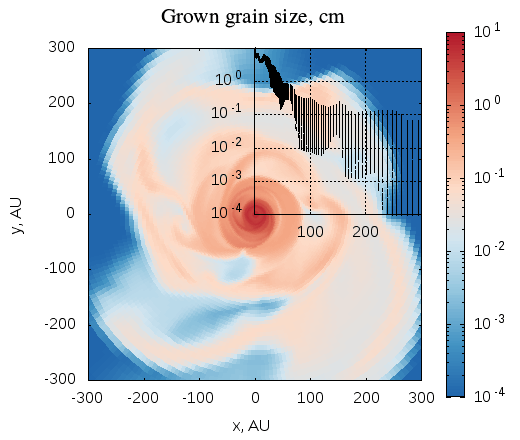}
\includegraphics[width=\columnwidth]{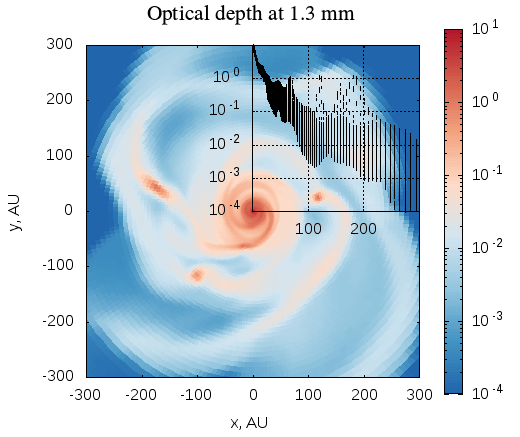}
\includegraphics[width=\columnwidth]{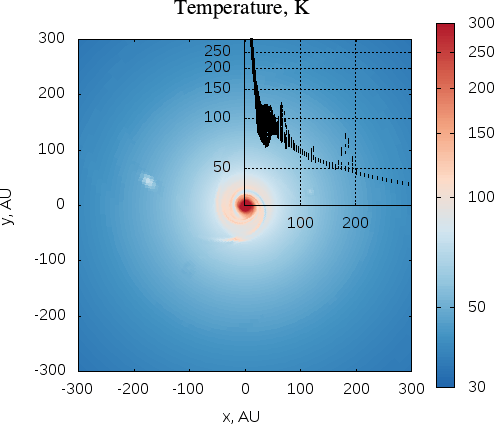}
%\fig{fig03-aa.png}{0.49\textwidth}{}
%\fig{fig03-bb.png}{0.49\textwidth}{}\\
%\fig{fig03-cc.png}{0.49\textwidth}{}
%\fig{fig03-dd.png}{0.49\textwidth}{}\\
\caption{
Circumstellar disc structures obtained with the {\tt FEOSAD} code at 0.16~Myr after formation of the central star. Top left: Map of gas surface density. Top right: Map of the maximum grown grain size $a_{\rm r}$. Bottom left: Map of optical depth at 1.3 mm. Bottom right: Map of temperature. The radial distributions of the same quantities for all azimuthal points are shown in the upper right corner of each map.}
\label{fig4}
\end{figure*}

In Fig.~\ref{fig4}, the distributions of gas surface density, the maximum radius of grown dust $a_r$, the total optical depth (determined by both grown and small dust grains) at 1.3 mm, and the temperature are shown.
At 0.16~Myr after the formation of a star, the disc shows a well-developed spiral structure with three clumps located between 100 and 200~AU, which are visible on the gas surface density and optical depth maps. The inner disc region at $r<20$~AU is optically thick ($\tau >1$), the clumps are moderately optically thick ($\tau \approx 1$), while the rest of the disc is optically thin ($\tau \ll 1$) at 1.3~mm. The maximum grown dust size varies from $10^{-4}$~cm at 300~AU to 10~cm at the disc centre. We note that the adopted {\tt FEOSAD} model does not predict the significant dust growth within the clumps, while there is an obvious correlation between the maximum grain size and the spiral features in the sense that regions of the maximum grain size clearly coincide with the spiral arms visible on the gas surface density map and the optical depth map.

\begin{figure}
\centering
\includegraphics[width=\columnwidth]{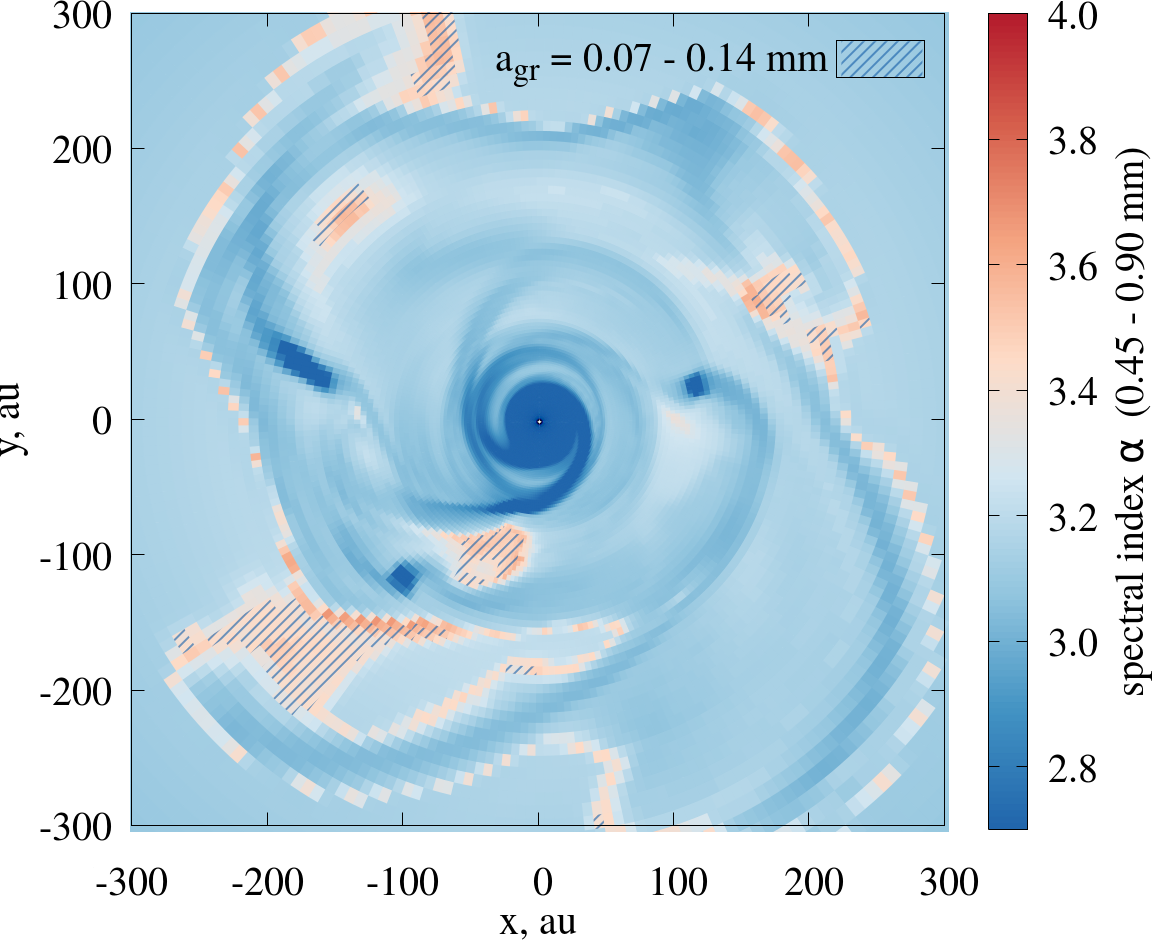}\\
\includegraphics[width=\columnwidth]{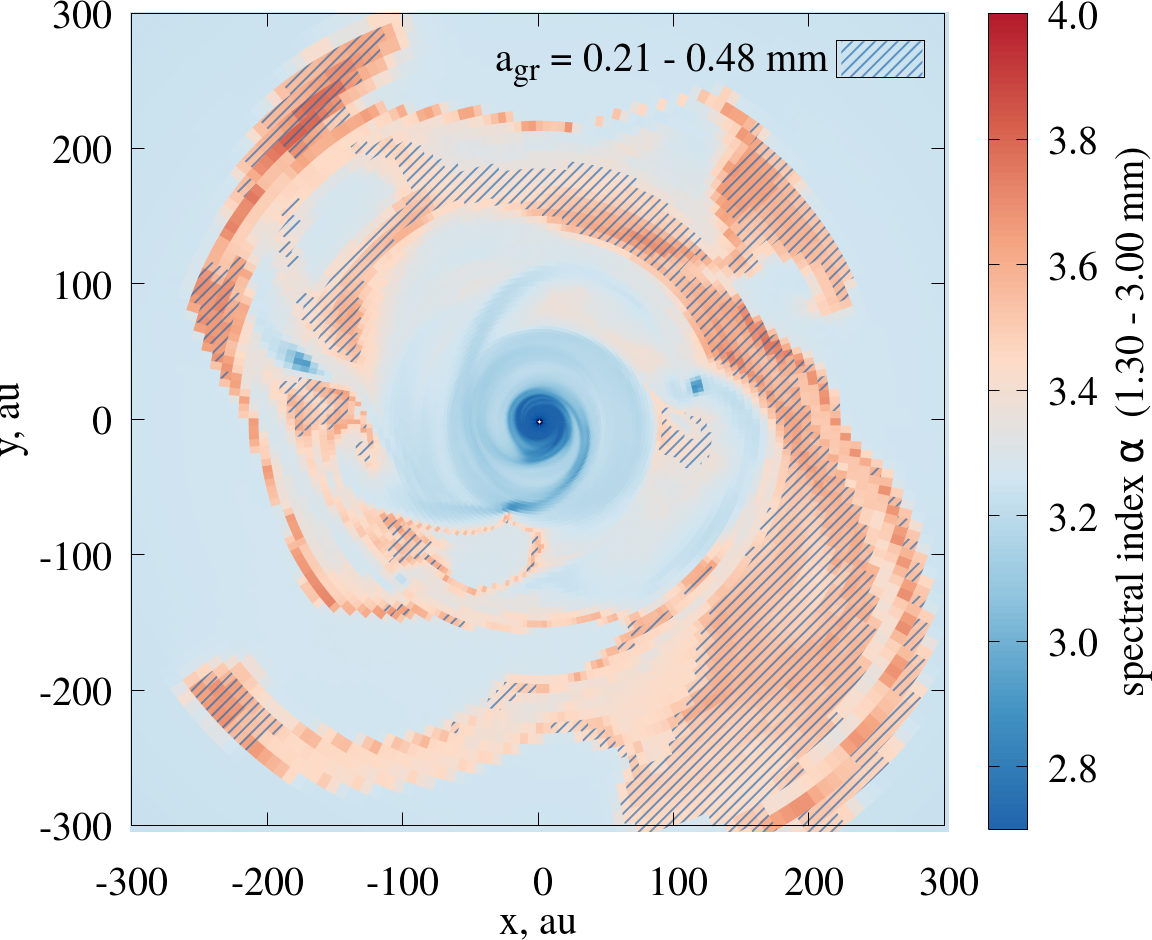}\\
\includegraphics[width=\columnwidth]{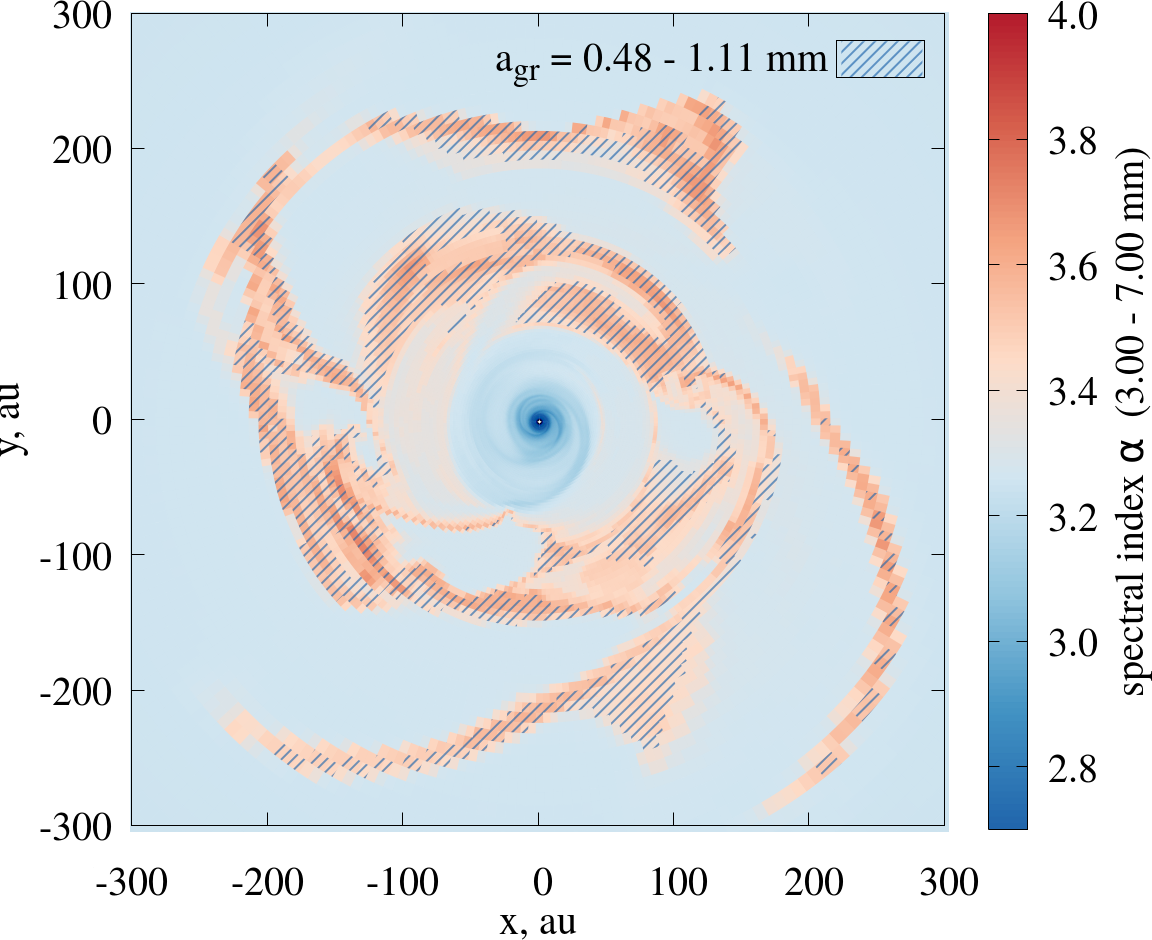}\\
%\fig{fig04aa.png}{0.4\textwidth}{}\\
%\fig{fig04bb.png}{0.4\textwidth}{}\\
%\fig{fig04cc.png}{0.4\textwidth}{}\\
\caption{Spatial maps of the spectral index $\alpha$ calculated between $0.45-0.9$\,mm (upper panel), $1.3-3.0$\,mm (middle panel), and $3.0-7.0$\,mm (bottom panel) for the model disc of 0.16~Myr age. The hatched area shows the location of dust grains with sizes  $\lambda_1/2\pi < a_{\rm gr} < \lambda_2/2\pi)$ corresponding to the maximum of the opacity index. Note the spatial correlation between the maximum in $\alpha$ and the locations of these grains.}
\label{fig5}
\end{figure}

\begin{figure}
\centering
\includegraphics[width=\columnwidth]{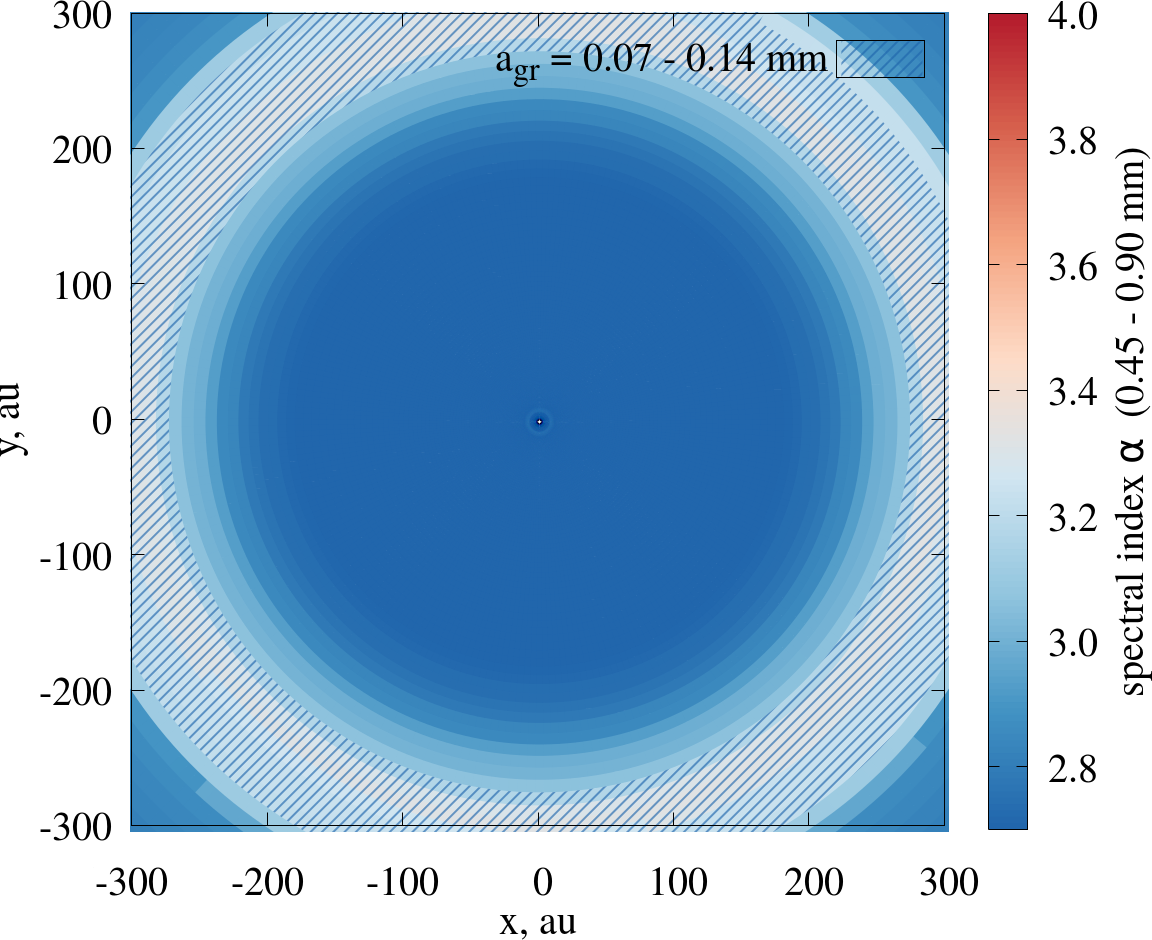}\\
\includegraphics[width=\columnwidth]{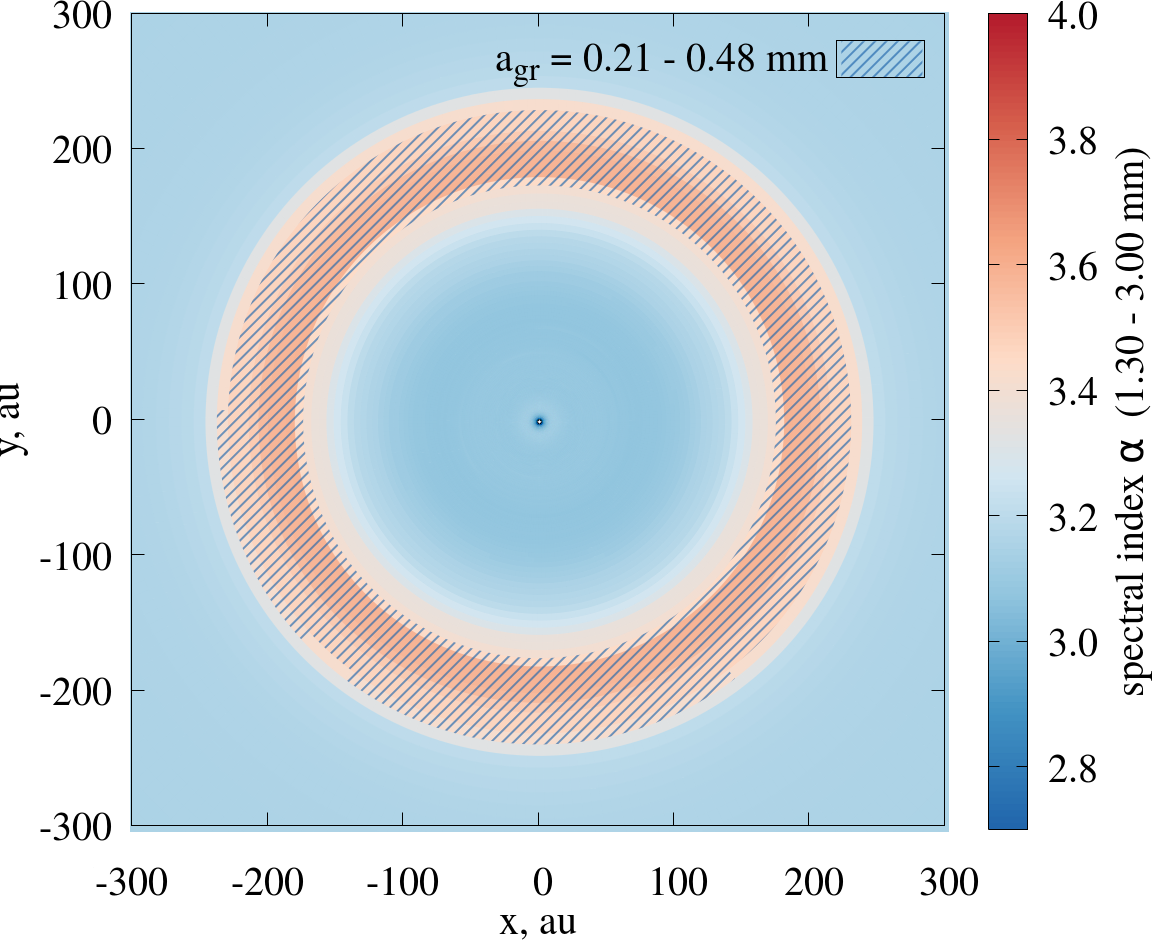}\\
\includegraphics[width=\columnwidth]{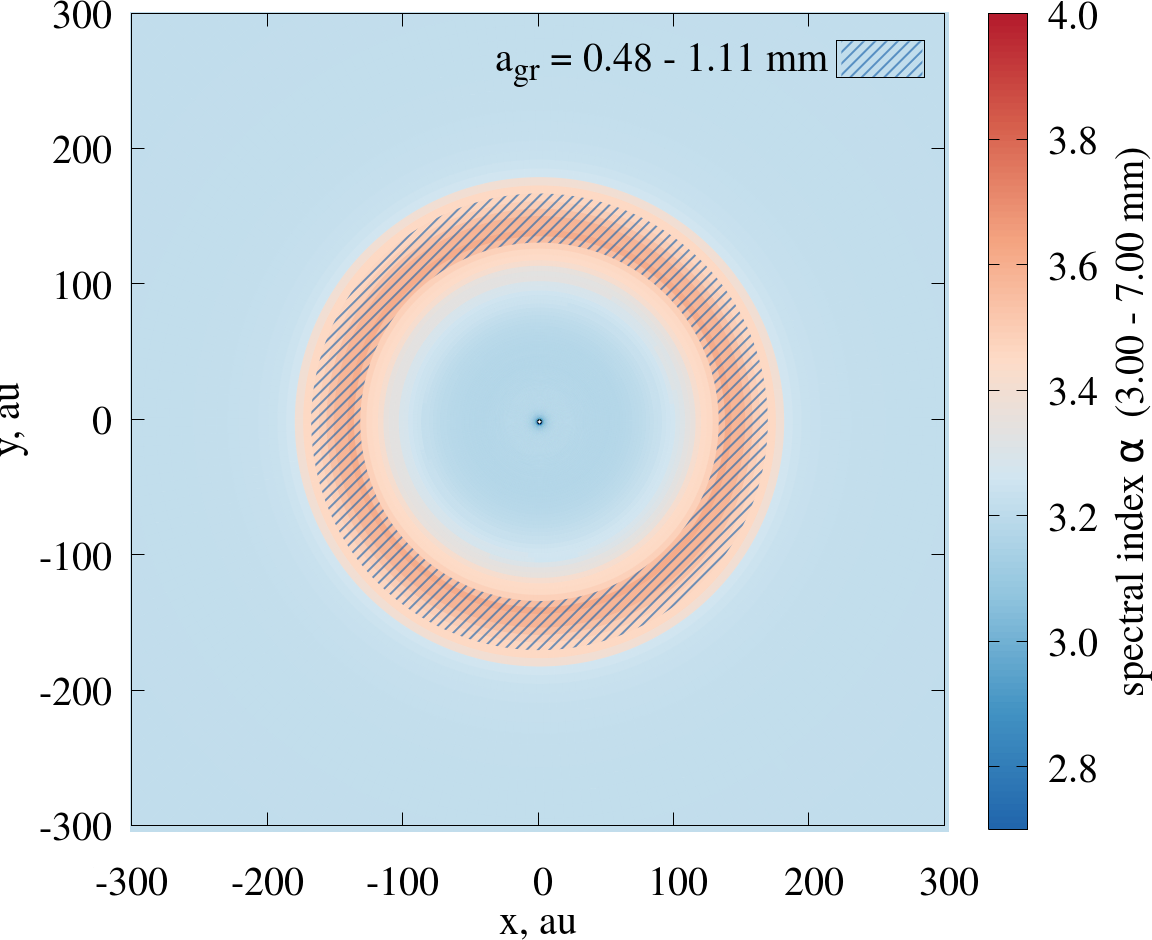}\\
%\fig{fig04aa-2.png}{0.4\textwidth}{}\\
%\fig{fig04bb-2.png}{0.4\textwidth}{}\\
%\fig{fig04cc-2.png}{0.4\textwidth}{}\\
\caption{The same as in Fig.~\ref{fig5} but for the model disc of 1~Myr age. 
\label{fig6}}
\end{figure}

The corresponding spectral index distributions calculated using the same three wavelength intervals as in Fig.~\ref{fig1} are shown as colour maps in Fig.~\ref{fig5}.
On all the three maps, the spectral index has minima toward the inner regions of the disc since the optical depth is high there. The optical depth becomes lower at larger wavelengths, therefore the size of the inner opaque region with low $\alpha$ is the largest at 0.45--0.9~mm (the smallest wavelengths considered in this study). By the same reason the clumps are clearly seen as ``blue'' knots at a 0.45--0.9~mm map but disappear at the 3.0--7.0~mm spectral index map. The key feature of the maps is the presence of regions with high values of the spectral index $\alpha$. These  regions shift closer to the disc centre as we switch to longer wavelengths. From the previous analysis it follows that these high-$\alpha$ regions should coincide with the spatial locations of grains having a specific maximum size. Namely, the maximum radii in the corresponding grain ensembles should be confined within the interval $\lambda_1/2\pi<a_{\rm gr}<\lambda_2/2\pi$, where $\lambda_1$ and $\lambda_2$ are the boundaries of the wavelength interval over which $\alpha$ is estimated (see Fig.~\ref{fig3}). Spatial locations of dust grains within these specified size intervals are shown as hatched regions in Fig.~\ref{fig5} for each of the considered wavebands. Obviously, for each $(\lambda_1,\lambda_2)$ interval there is a strong correlation between regions of maximum $\alpha$ in this interval and location of dust grains with sizes within the $(\lambda_1/2\pi,\lambda_2/2\pi)$ interval. Our idea is to use this correlation as a tool to associate peaks on the observed spectral index maps with the locations of grains having specific values of $a_{\rm max}$, which in turn are defined by the wavelength interval used for the evaluation the spectral index.  When a single spectral index map is available, this method does not provide the information about $a_{\rm max}$ in other parts of the map, except for those toward the peaks of $\alpha$. Therefore, one has to use multi-frequency $\alpha$-maps derived using several wavelength intervals to infer the spatial localization of dust with different values of $a_{\rm max}$.

We note that the presented maps are based on the particular dynamical disc model and correspond to very early stages of the disc evolution, $t=0.16$~Myr. This age was selected since the disc exhibits a complex structure (due to the gravitational instability) and thus provides a good benchmark for the method. However, we expect to see the same correlation for later stages and for other dynamical disc models as long as we rely on the dust opacity dominated by the spherical solid dust grains. As an illustration, in Fig.~\ref{fig6} we present the spectral index maps for the same dynamical disc model but after 1~Myr of evolution. Obviously, the morphology of the maps became more regular and axisymmetric than at 0.16~Myr but still there is a strong correlation between high values of the spectral index and the maximum grain sizes.

\begin{figure*}
\centering
\includegraphics[width=0.66\columnwidth]{fig04bb.png}
\includegraphics[width=0.66\columnwidth]{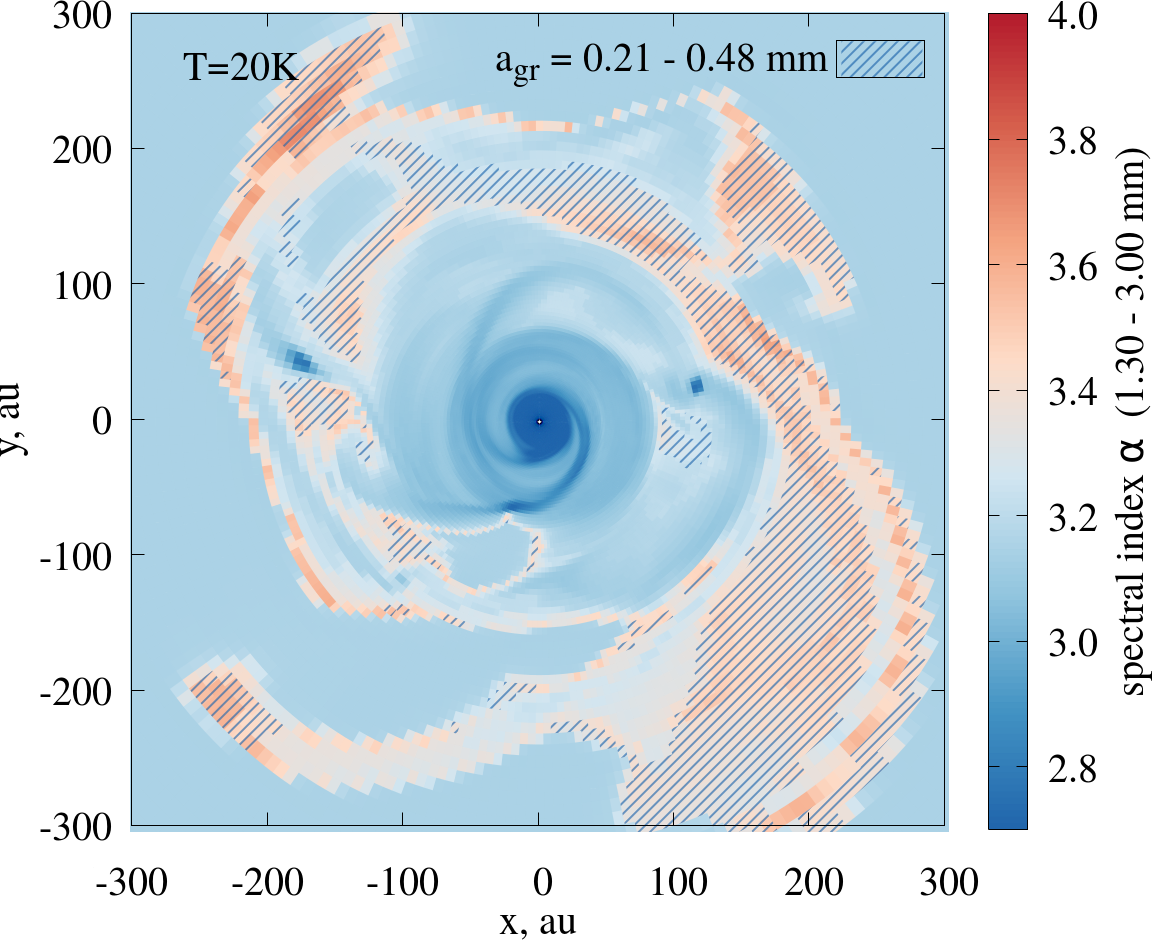}
\includegraphics[width=0.66\columnwidth]{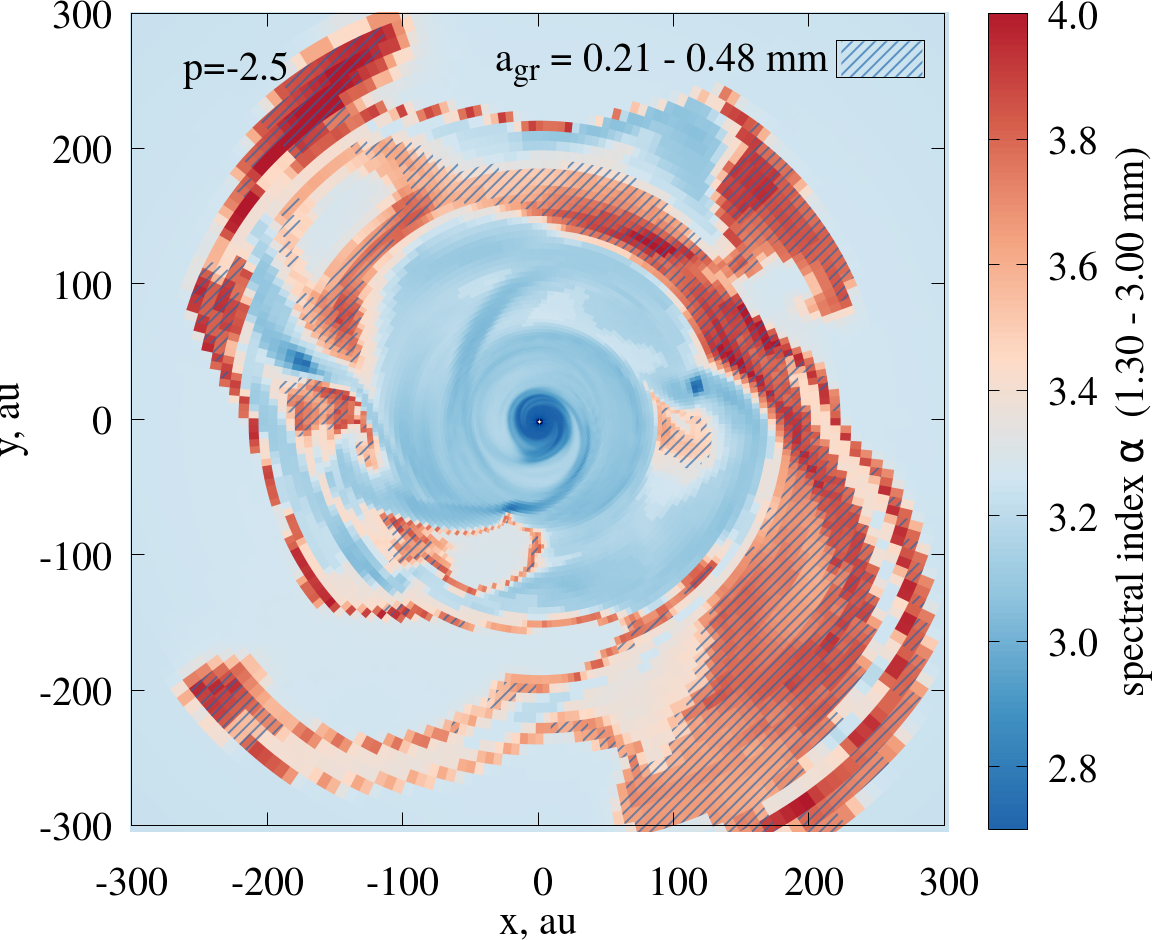}
\caption{Spatial maps of the spectral index $\alpha$ calculated in a wavelength interval of $1.3-3.0$\,mm for the original disc model of 0.16~Myr age (left); for the model where the temperature is manually set to $T=20$~K (middle); for the model where the power law slope of dust size distribution is manually set to $p=-2.5$ (right). The hatched area shows the location of dust grains with sizes  $\lambda_1/2\pi < a_{\rm gr} < \lambda_2/2\pi$, where $\lambda_1=1.3$~mm and $\lambda_2=3.0$ mm.}
\label{fig7}
\end{figure*}

One may expect that the local maxima in $\alpha$ can be caused by other factors, e.g. by temperature gradients. Indeed, the temperature can affect $\alpha$ if $x=h\nu/kT \gtrsim 1$, see Eq.~\eqref{eq:main}. In principle, strong and non-monotonic temperature gradients in cold parts of the disc could produce maxima in the spectral index maps. At the same time, the location of the $\alpha$-maximum emerged due to the maximum of $\beta$ is unlikely to be affected by the details of underlying temperature distribution. To demonstrate this, we present an additional spectral index map in Fig.~\ref{fig7} (middle panel). This map is calculated using a constant temperature of $T=20$~K rather than the original temperature distribution ranging from 20~K to 400~K in the dynamical model of the disc. As a result, the values of spectral index indeed change but the maxima in $\alpha$ are still present at the same positions, and the correlation is unaffected.

To produce the first three charts in Fig.~\ref{fig3} and the spectral index maps in Figs.~\ref{fig5}--\ref{fig6}, we used $a_{\rm min}=1$~$\mu$m and a grain size distribution slope of $p=-3.5$. However, the actual dust size distribution in protoplanetary discs may be different. In particular, the chosen $a_{\rm min}$ is larger than its value in the typical interstellar medium (MRN-like), which is in the range 0.005 $\mu$m < $a_{\rm min}$ < 0.25~$\mu$m, see~\citet{1977ApJ...217..425M}. We note, however, that the $\beta$ index does not strongly depend on $a_{\rm min}$ as long as $2\pi a_{\rm min}/{\lambda} \ll 1$. The choice of the power-law slope $p=-3.5$ implies that there are many grains on the small size end of the dust size distribution. A flatter slope would produce a distribution with more grains of the larger size, and is probably better suited to model the protoplanetary disc midplane where most of the sub-mm/mm emission is produced. Therefore, we additionally checked that the selections of a shallower power law do not influence the location of the spectral index peaks. As an example, in Fig.~\ref{fig7} (right panel) we show the spectral index map for the model where the exponent of the dust size distribution is manually set to $p=-2.5$, keeping all the other distributions as in the original model. It is seen that values of $\alpha$ became significantly larger but the morphology of the map remained the same.

Thus, we suggest the following simple method to study dust segregation in protoplanetary discs. If intensity maps for a certain disc are available in several wavelength intervals (e.g., in the ALMA bands), one can plot maps of $\alpha$ for each of these wavelength intervals, and regions of maximum $\alpha$ will indicate the spatial locations of dust grains of the corresponding maximum sizes.

Finally, it is necessary to comment on the validity of the method given possible uncertainties when deriving $\alpha$ from observed maps. The uncertainty in $\alpha$ can be caused by atmospheric noise or, more significantly, by the uncertainty on the absolute flux calibration. In the first case, we expect our method to work if the noise $\sigma_{\alpha}$ in the measured spectral index is significantly lower than the peak strength in the spectral index map, i.e. when $\sigma_{\alpha} < \alpha_{\rm peak} - \alpha_{\rm bg}$, where $\alpha_{\rm peak}$ is the peak value of the spectral index, and $\alpha_{\rm bg}$ is the value of the spectral index in the immediate vicinity of the peak. 
In the case of poor flux calibration we expect that the inferred values of $\alpha$ will deviate from the true values but the spatial location of spectral index peaks should not be affected. However, one needs to make a more detailed analysis how to account for these uncertainties when using the proposed method.

\section{Conclusions} \label{sec:conclusions}

As grain sizes in protoplanetary discs are expected to be of the order of the observational wavelength in far-IR, the commonly used approximation on the grain opacity $\kappa\propto \nu^{\beta}$ with a fixed opacity index $\beta$ may not be appropriate. The value of $\beta$ depends on the chosen wavelength interval and grain population (see Fig.~\ref{fig2}). In the case of spherical solid dust grains, the presence of large amounts of dust grains having size $a$ considerably increases the $\beta$-value around a wavelength $\lambda\approx2\pi a$. 

In the optically thin case, the opacity index $\beta$ is traced by the SED spectral index $\alpha$ (see Eq.~\ref{eq:main}). This means that the spatial distribution of the \textit{maximum} of the spectral index $\alpha$ should correlate with the spatial distribution of grains with sizes corresponding to the wavelength  interval where $\alpha$ was determined. We demonstrate this correlation using the theoretical modeling of (bi-disperse) dust evolution coupled with hydrodynamical simulations of a gravitationally unstable disc (see Fig.~\ref{fig5}).

The important aspect is the use of well calibrated observational data with a sufficiently high signal-to-noise ratio which allows deriving reliable spectral index maps. While selecting the wavelength intervals one has to find a compromise between intention and capability. On the one hand, the position and width of the wavelength interval should be related to the expected ranges of the maximum grain sizes. On the other hand, the observational data should be good enough to derive this information. The obvious direction to rectify this method is to study the correlation between the maximum grain size and maximum $\alpha$ for other possible grain size distributions and compositions.

%\acknowledgments 
We thank the referee for the very relevant comments and constructive suggestions. The study was supported by the Russian Scientific Foundation grant 17-12-01168.

%%%%%%%%%%%%%%%%%%%%%%%%%%%%%%%%%%%%%%%%%%%%%%%%%%



\bibliographystyle{mnras}
\bibliography{reference} % if your bibtex file is called example.bib

%\begin{thebibliography}{99}
%\end{thebibliography}

%%%%%%%%%%%%%%%%%%%%%%%%%%%%%%%%%%%%%%%%%%%%%%%%%%

%%%%%%%%%%%%%%%%% APPENDICES %%%%%%%%%%%%%%%%%%%%%

%\appendix

%%%%%%%%%%%%%%%%%%%%%%%%%%%%%%%%%%%%%%%%%%%%%%%%%%

% Don't change these lines
\bsp	% typesetting comment
\label{lastpage}
\end{document}